\documentstyle[aps,epsfig]{revtex}
\newcommand{\be}{\begin{equation}}
\newcommand{\ee}{\end{equation}}

\begin{document}
\draft
\title{Growth Exponents with 3.99 Walkers}
\author{M. B. Hastings}
\address{
CNLS, MS B258, Los Alamos National
Laboratory, Los Alamos, NM 87545, hastings@cnls.lanl.gov 
}
\date{17 April 2001}
\maketitle
\begin{abstract}
It is argued that the dielectric-breakdown model has an upper critical
$\eta_c$ equal to 4, for which the clusters become one-dimensional.  A
renormalization group treatment of the model is presented near the
critical $\eta$.
\vskip2mm
\end{abstract}
\section{Introduction}
The model of diffusion-limited aggregation (DLA)\cite{dla} has presented
a great challenge to theorists.  The model describes many phenomena, including
viscous fingering\cite{vf}, electrodeposition\cite{dep}, and dendritic
growth\cite{dend}, but has also become important as a theoretical problem
of its own.  While the growth rules are simple, they are highly nonlocal and
give rise to complex branching structures that cannot be described easily
by any small perturbation of a smooth surface.

Despite much important recent theoretical work on the structure of 
DLA\cite{thy,thy2,thy3}, these attempts have all involved approximations
or phenomenological assumptions, without a fully controlled expansion.
In analogy to critical phenomena, we would also like an $\epsilon$ expansion
to provide a framework which may be systematically improved.

It is believed that DLA remains nontrivial in any finite dimension, so
an expansion about an upper critical dimension is not possible\cite{ft}.  We
turn instead to another generalization of DLA, the dielectric-breakdown
model(DBM)\cite{dbm}.  This model offers a continuously varying fractal
dimension as a function of a parameter $\eta$, ranging from 2 at $\eta=0$ to
approximately 1.7 at $\eta=1$ (DLA) to 1 at $\eta=\infty$.  The model is
equivalent to simultaneously releasing $\eta$ random walkers and requiring
that they all hit a given point for growth to occur.  As the model remains
nontrivial for $\eta\rightarrow 0^+$ \cite{eta0}, we seek instead an
expansion about an upper critical $\eta_c$, for which the clusters
become one-dimensional.  One attractive feature of this approach will
be that near $\eta_c$ the clusters are described in terms of one-dimensional
branches, so that the branching structure of DLA is inherent.

Previous numerical studies have suggested $\eta_c=4$ \cite{num1}, as well
as providing some analytic explanation.  A recent study of much larger
clusters\cite{num2} also indicates a finite $\eta_c$ between 4 and 5.
This study found significant finite size corrections for clusters with
$\eta \geq 4$, making it problematic to determine dimension with small
clusters.

In this work we provide an analytic argument for $\eta_c=4$ based
on branch competition, and present the lowest order in a $4-\eta$ expansion.
The techniques are related to the branched growth model\cite{thy2} and to
singular Laplacian growth\cite{slg}, while the way we evaluate growth
processes in the long-time limit is close to the fixed-scale 
transformation\cite{thy3}.

We proceed as follows: first we argue the equivalence, at least near $\eta=4$,
of the DBM to a reduced model based on one-dimensional branches with
discrete tip-splitting events.  Next, we analyze the competition of two
branches and show that on large scales tip-splitting events proliferate
for $\eta<4$, while are suppressed for $\eta>4$.  Then we consider the
competition of three branches (using numerical and analytical techniques)
and show that these higher order effects suppress the tip-splitting, leading
to an attractive fixed point at a finite value of the tip-splitting rate,
with no relevant perturbations of the fixed point.  We then use this fixed
point to determine fractal and multifractal dimensions and compare to numerics.

The renormalization group used involves expanding simultaneously
in $4-\eta$ and the tip splitting rate.  A related tip-splitting expansion 
to systematically extend the branched growth model has been suggested 
previously\cite{unpub}.  We consider a system with
given ratio between microscopic and macroscopic cutoffs and evaluate the
probability of various growth processes, to obtain corrections to the
growth rate of the cluster and to the tip-splitting rate; we find that
all such quantities can be written as a sum of logarithmic corrections
to bare quantities.  We then make an assumption that the logarithms
may be resummed to produce power laws.  This assumption relies on the
renormalizability of the model; we have no proof of renormalizability, but
in the last section we sketch how such a proof might proceed. 
Resumation of logarithms has been used for other nonequilibrium systems
such as Barenblatt's equation\cite{ng}
\section{Reduced Model}
Since we wish to have an RG in which power laws are obtained by resummation
of logarithms, it is essential to remove all irrelevant variables from
the problem.  We construct such a model in two steps: first we go from
the conformal mapping model for DLA to a discrete version of the model.
Next, we go to a reduced version of the model in which branches have
vanishing aspect ratio and grow deterministically.  The discrete model
will be useful in providing a definition of tip-splitting events, which
will provide the only source of randomness in the reduced model.

Recall the conformal mapping formulation\cite{conf} of DLA and the 
dielectric breakdown model: consider a function $F^{(n)}(z)$ that maps a
straight line onto
the boundary of the growing cluster after $n$ growth steps.  To obtain 
$F^{n+1}$, pick a point $w$ with probability $|F^{(n)'}(z)|^{1-\eta}$.  Define
an elementary mapping $f^{(n+1)}(z)$ that produces a bump of linear size
$\sqrt{\lambda}$ at $w$, where, to obtain the correct bump size in the physical
plane,
\be
\lambda=|F^{(n)'}(w)|^{-2}.
\ee
Then, define $F^{(n+1)}(z)=F^{(n)}(f^{(n+1)}(z))$.

Next, we consider a discrete version of the above model.  Each elementary
mapping $f$ has square-root singularities located a distance of order
$\sqrt{\lambda}$ from $w$.  All points $z$ between these singularities are
mapped onto the surface of a given bump (though, after further growth
steps, some of these points may be mapped onto bumps growing off the given 
bump).  In the discrete model, we will first pick a point $w$ as above; then,
we find the nearest square-root singularities on each side of $w$, and grow
a bump at a point equidistant between those singularities.  This simple
modification means that there are now a discrete set of growth sites (in
any lattice formulation of the problem, one would also find a discrete set
of growth sites).

Now, assume that the cluster has a roughly one-dimensional shape, with a tip 
near $z=z_i$ (we reserve an index $i$ to label different such tips).  
If the branches were precisely one-dimensional, near
the tips the singular behavior of $F(z)$ would be given by
\be
\label{sing}
F(z)=F_i (z-z_i)^2,
\ee
with $F_i$ some constant.  Due to the finite size of particles in the model,
$F(z)$ is given by a suitably regularized version of Eq. (\ref{sing})
near the tip.  For $\eta>2$, the
integral of the measure $|F'(z)|^{1-\eta}$ diverges for small $z$, so that
in the discrete 
model, the probability measure will be concentrated on the one growth site at 
the tip, with rapidly decaying measure on the neighboring sites.  The measure
on neighboring sites will be nonuniversal and determined by the particular
form of elementary mapping $f$ we choose, which will define a particular
regularization of the model.  

Consider a branch which grows without splitting.  While the length of
the branch increases constantly in time, the width remains of order the
microscopic scale.  Thus, at large scales such a branch looks like
a singular configuration with vanishing aspect ratio.  At the
upper critical $\eta$, the assumption of growth without splitting will be valid
at large scales,
while below the upper critical $\eta$ the clusters are approximately
described by a collection of one-dimensional branches,
implying that the aspect ratio is an irrelevant variable near the
upper critical $\eta$.
So, in constructing the reduced model, we will take all branches 
with vanishing aspect ratio.

As the aspect ratio vanishes,
all of the growth measure becomes concentrated
near the growth tip, on a scale much less than the length of the branch.
However, even though the aspect ratio vanishes,
due to the discretization there is some growth measure
on sites neighboring the growth tip, and growth may arise on any of these 
sites, with some nonuniversal probability.  This is a tip-splitting event.
Below, we will assign a tip-splitting rate, and consider how this
rate renormalizes.  As we expand to higher orders in tip-splitting,
we include more growth configurations, so that to sufficiently high order
we restore the full set of clusters found in the original model.

We now determine the growth measure for a branch, based on the proper
regularization of Eq. (\ref{sing}).  The correct microscopic
regularization of the model is at fixed length in the physical plane, leading 
to a cutoff
at length of order $1/\sqrt{|F'|}$ in the $z$-plane.  Thus, the correctly
regularized value of $F'$ at the growth tip is 
\be
\label{regfp}
\sqrt{F_i}.
\ee

Numerically, one may implement this by evaluating $F'$ a short distance
from $z$.  We note that the difference in $F'$ depending on 
regularization means that different short distance regularizations may
imply completely different physical models; this may be behind some
of the differences found between diffusion-limited aggregation and
Laplacian growth\cite{clg}.

Eq. (\ref{regfp}) is different from that taken in the idea of singular
Laplacian growth\cite{slg}, and reflects the correct regularization of the
problem in the physical plane.
Similarly, the correctly regularized growth measure at the tip, which in the
discrete model is an {\it integral} of $|F'|^{1-\eta} {\rm d}z$ over the 
region between singularities, is
\be
\label{reggm}
v_i=\frac{\sqrt{F_i}^{-\eta}}
{\sum\limits_i\sqrt{F_i}^{-\eta}}.
\ee
Further, we assume in constructing the reduced model
that if there is a collection of growth sites, at large
scales it is valid to grow each site deterministically
at a velocity proportional to
Eq. \ref{reggm}, ignoring the shot noise inherent in the discrete formulation
of the model.  As the branches becomes long compared to the walker scale, this
assumption is valid.  All the randomness in the reduced model will arise via
tip splitting.

In what direction does a tip grow?  In our model, each microscopic
growth step is produced by a mapping $f$, growing the surface in the
$z$ plane normal to itself at $z=z_i$.  For a nonsingular configuration, this
grows the surface in the physical plane normal to itself; however, the tip
of a one-dimensional configuration has no normal direction, and we are
forced to rely on the conformal mapping definition of growth.  In
the conformal mapping model, growth at $z_i$ for time ${\rm d}t$ with
velocity $v_i$ can
be obtained by composing $F$ with a function 
$f=z+v_i{\rm d}t\frac{\lambda}{z_i-z}$.
As growth progresses, the singularities $z_j$ move, as the
point which is mapped to a singularity $j$ of $F$ is $f^{-1}(z_j)=z_j
-v_i{\rm d}t\frac{\lambda}{z_i-z_j}$.
This leads to
\be
\label{cnd}
\partial_t z_i=\sum\limits_{j\neq i}
\frac{1}{z_i-z_j}
F_j^{-1} v_j.
\ee

Our model of growth causes a tip $z_i$ to grow in the direction $F_i$
in the complex plane, but as $F_i$ may change in time, the tips may
curve.  This differs from the model of singular Laplacian growth in which 
branches always grow in a straight line, and in which there are additional terms
in the motion of each tip due to growth at that tip.  
However, with correct regularization in which the map function $f$ is
chosen to produce a bump of small but finite size, one may show that
$z_i$ does not move due to growth at tip $i$ itself.  

If branches are straight, then $F(z)$ has a finite number of singularities
at branch points and growth tips.  Any singularity, $z$, of $F$, which
is not a growth tip, obeys an equation of motion
\be
\label{sngd}
\partial_t z=\sum\limits_{i}
\frac{1}{z-z_i}
F_i^{-1} v_i.
\ee
If the branches curve, additional singularities are continuously produced.

Thus, we present the reduced model: let the cluster be described by
a mapping $F(z)$ which produces a cluster made of one-dimensional branches,
with growth tips labelled by $i$.  Grow each tip at a velocity
given by the growth measure of Eq. \ref{reggm} multiplied by a
growth velocity $v$.

In addition, tips split.  Assign a rate $g$, which is an
appropriately defined rate at which each tip has a tip splitting event;
the definition of $g$ is to a certain extent arbitrary and will
be discussed more below.  This rate is measured with respect to
the time integral of the growth probability measure on the given tip, so
that tips with greater growth probability measure split more readily.
The initial conditions for the two daughter branches are randomly chosen,
as one or another daughter may have the majority of the growth probability.
A regularization is necessary; as we show below, if the two daughter tips are 
created an infinitesimal distance apart with a finite asymmetry in
initial conditions, one tip will immediately dominate
the growth over the other.  Thus, we will need to assign a short distance
cutoff, so that the two tips survive for some minimum time, or to
some minimum finite separation.
In the RG we will use a particular cutoff; however, the precise form
of the cutoff will not be important.
\section{Geometry}
Throughout, we will work in cylindrical geometry in the long-time regime.
The size of the individual particles in the original model defines a microscopic
cutoff length, $l$, which will provide the minimum distance at which
branch pairs are created.  The horizontal width of the cylinder in which the
cluster grows define a macroscopic length scale $L$.  These cutoffs in 
length are related to cutoffs in time for branches, such that the
minimum time a branch exists is of order $l/v$.
In cylindrical geometry, the ratio between these 
cutoffs is fixed and the front grows upward at constant 
velocity.  Below we will expand in the number of tip splitting events,
assuming that the cluster is evolved for a finite time $T$, so that the
average number of tip splittings,
$g T$, serves an expansion parameter; $T$ must be taken infinite at the end
to obtain the long-time regime.

In radial geometry, the ratio between cutoffs is constantly
changing, leading to discrepancies between different measures of the
dimension\cite{diffd}.  An interesting future problem will be to understand
radial geometry in our framework, or to deal with the affine regime in
cylindrical geometry\cite{evert}.  

In cylindrical geometry, the intersection of
the cluster with a line at constant height is a fractal with dimension $D$
between 0 and 1.  The cluster grows upwards with velocity $v_{\rm ren}$.  
As the area of the cluster increases at a constant rate in time,
$v_{\rm ren}$ is related to the ratio of scales and the
microscopic velocity $v$ by
\be
\label{dim}
v_{\rm ren}\propto v (l/L)^D.
\ee
Later, we will compare the
analytic results to numerics on the DBM in radial geometry, so
we must assume that any difference due to geometry is small.  This difference
is small for DLA, roughly the difference between 1.66 and 1.71
and we assume that we may add 1 to the dimension given by
Eq. \ref{dim} to obtain the dimension in radial geometry.
\section{Competition of Two Branches}
Consider two competing branches, the first order in $g$ in the reduced model.
We linearize near a symmetric configurations of the two
branches.  The branch shape depends on initial conditions.  However,
in the long time limit, two symmetrically competing branches will grow straight.
Physically, the angle between the two branches cannot be too narrow due to 
the mutual screening effects; neither cannot be too large or the branches
would end up growing back towards their parent.  We will compute this angle
below.  If initially the branches are not at the given angle, they will
curve and approach this angle; we have observed this curving in numerical
simulations of the reduced model.

\begin{figure}[!t]
\begin{center}
\leavevmode
\epsfig{figure=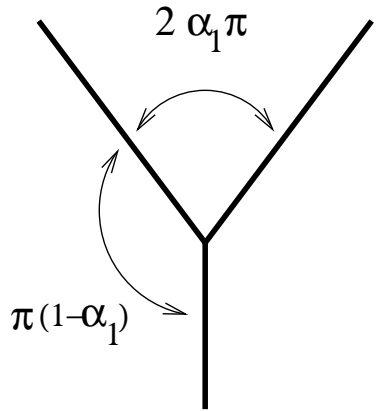,height=8cm,angle=0,scale=.75}
\end{center}
\caption{Competition of two branches.  Opening angles are indicated in
terms of $\alpha_1$.}
\label{fig1}
\end{figure}

If the branches are straight in the
physical plane, $F$ is a degenerate Schwarz-Christoffel map.
We define $F(z)$ by its derivative:
\be
\label{feq}
F'(z)=(z-y)^{-\alpha_1} (z+y)^{-\alpha_1}
z^{-\alpha_2} (z-x) (z+x),
\ee
describing growth tips are $z=\pm x$.  We have $2 \alpha_1 +\alpha_2=1$.
In Fig. \ref{fig1} we show the resulting configuration and relate the
angles to $\alpha_1,\alpha_2$.

We will use a trick to describe the dynamics of this system.  The
reduced model above has a specified dynamics for the motion of the
growth tip singularities $x$.  The model of singular Laplacian growth,
a different model, which admits solutions with straight branches
of {\it arbitrary} opening angle, has
a different set of solutions for the motion of $x$.  The equations
of motion of the two models match only when 
\be
\label{cndn}
\partial_z \Bigl(\frac{F'(z)}{z\pm x}\Bigr)=0,
\ee
 for $z=\mp x$.
This use of singular Laplacian growth is simply a trick to derive the
constraint (\ref{cndn}) that
a Schwarz-Christoffel map must obey to grow straight under the
reduced model dynamics.

If the map remains Schwarz-Christoffel, we use Eq. (\ref{cnd}) to find the
motion of $x$, as well as Eq. (\ref{sngd}) to describe the motion of $y$.
We find:
\be
\label{xd}
\partial_t x=\frac{1}{2}\frac{1}{2x},
\ee
\be
\partial_t y=\frac{1}{2}\Bigl(\frac{1}{y-x}+\frac{1}{y+x}\Bigr),
\ee
where we normalize $F_i=1$ for the two different tips $i=1,2$.
We note that for this configuration $F_i$ is constant in time.

Requiring that $\partial_t y/\partial_t x=y/x$, we find that
\be
y=\sqrt{5} x.
\ee
Requiring Eq. (\ref{cndn}), to obtain straight growth,
we find
\be
\alpha_1=1/5,\alpha_2=3/5,
\ee
which implies that the competing branches have a 72 degree opening angle.
In numerical simulations with $\eta$ near 4\cite{num2}, this characteristic 
opening angle can be clearly seen.

Having found $F$ when the two branches are symmetric, we
now consider, to linear order, to competition of two asymmetric branches.
Let the branches have growth measures $v_1,v_2$.  Define 
$\delta=\ln{F_1/F_2}$.
We will find that the dynamics is unstable, and one of the two branches
will win; as that branch wins, it will curve, until it becomes parallel
to its parent.  However, to linear order, we may ignore the curvature, and 
assume that the map retains the Schwarz-Christoffel form.  Above,
we noted that the branches grow in direction $F_i$; while the
competition moves the singularities $z_i$, leading to changes in $F_i$ and 
curvature of the branches, Eq. (\ref{cndn}) implies that to {\it linear}
order the motion of singularities does not change the angle at which the
branches grow.


Eq. \ref{feq} generalizes to
\be
\label{gfeq}
F'(z)=(z-y+\delta_1)^{-\alpha_1} (z+y+\delta_2)^{-\alpha_1}
(z+\delta_3)^{-\alpha_2} (z-x+\delta_4) (z+x+\delta_5).
\ee
Assume the various $\delta$ are all 
small.  
Assuming that the map retains this form, one can
use the equations of motion for all five singularities to
determine the dynamics of the map.  Regardless of whether $v_i$ is
chosen from Eq. (\ref{reggm}) or chosen arbitrarily, Eqs. (\ref{cnd},\ref{sngd})
lead to constraints on the possible resulting $\delta$
which reflect constraints following from
the assumption that the Schwarz-Christoffel map is degenerate.  As a
result, the competition of two branches starting from
an almost symmetric configuration
occurs in a two-dimensional parameter space, specified by $\delta$ and
by the overall length scale of the branches.

After some algebra, one finds
$\partial_t \delta=(v_1-v_2)/t-\frac{\delta}{t}$, or
\be
\label{ddeq}
\partial_t \delta=\Bigl(\frac{\eta}{2}-1\Bigr) \frac{\delta}{t}.
\ee

The factor of $1/t$ arises from 
Eq. \ref{xd}, giving $x=\sqrt{t/2}$.
This factor implies
\be 
\label{pgr}
\delta \propto t^{\frac{\eta}{2}-1}.
\ee

The factor $t^{\eta/2}$ in the above equation reflects the branch competition. 
The factor $t^{-1}$, present even at $\eta=0$, reflects the fact that if both
branches grow at constant velocity, the difference in the lengths of the
two branches remains constant while the total length of each branch and
and the separation between branches
increase linearly in $t$, so that the difference in length scaled by
total length behaves as $t^{-1}$.

For $\eta=4$, $\delta$ grows linearly in time.  For large enough $\delta$,
nonlinear effects take over and one branch dies.  Thus, assuming the
branches are created after a tip splitting event with a probability that
is not singular at $\delta=0$, the probability of both branches surviving
for time $t$ is proportional to $1/t$ at $\eta=4$.  On the other hand, the
number of possible tip splitting events in time $t$ is proportional to $t$.
Thus, the probability of creating a branch in time $t$ that
survives for time $t$ is independent of $t$. 
This will become more clear in the next sections when we consider interaction
of three branches.  For $\eta>4$, this probability
decays for large $t$; on the other hand, for $\eta<4$, this probability
increases, and tip-splitting events proliferate.  Defining $\epsilon=4-\eta$,
the lowest order $\beta$ function is
\be
\label{bfn0}
\beta(g)=\frac{\epsilon}{2} g.
\ee
\section{Renormalization Group}
The renormalization group is based on simultaneously expanding in
$g$ and $4-\eta$.  In this section we outline how the RG works; in the
next section we will discuss the implementation of the RG.

We define
survival of two subbranches by requiring that neither
of the two subbranches has more than some given percentage of the
the total growth probability.   Then we define $g$ as the probability in
time $t$ that two branches are produced which survive for time $t$, which
relates the arbitrariness in survival to an arbitrariness in the
renormalization scheme.  Alternately define $g$ as the probability in
time $\tilde T$ that two branches are produced which survive up to a separation
$\tilde L \propto v \tilde T$; 
this second definition will be more convenient and will be used
in the numerical work below.  The scale $\tilde L$ is intermediate between the
scales $l,L$.

We will compute two RG functions, the renormalization of the growth
velocity and the renormalization of the tip splitting rate $g$.  Consider
the renormalization of growth velocity to first order in $g$ at
$\eta=4$.  If there is
only one branch in the system, this branch grows upwards with velocity $v$.
If the branch splits into two subbranches, the growth velocity is reduced until
eventually one of the two subbranches wins and the original growth velocity
is restored at long times, where no further tip splitting events intervene so
long as we work to first order in $g$.
The probability that the subbranches survive to distance $\tilde L$ scales as 
$1/\tilde L$.  

The probability that both branches survive exactly time $t$ and
distance $v t$ scales as $1/t^2$.
If the two branches survive for time $t$, the height of the cluster at
long times is reduced by an amount of order $v t$ compared to the height
of the cluster without tip splitting.  More precisely, evolve the
cluster for total time $T$, so that the probability of a single such
tip splitting event is $g T/t^2$, and compute the
average height $h$ of the cluster after time $T$, to find
\be
h=v T - g T c \int \frac{dt}{t},
\ee
where $c$ is some constant.  
This yields the renormalized velocity
\be
\label{lgh}
v_{\rm ren}=v 
(1-\frac{g c}{v} \ln{\tilde L/l}).
\ee
Interpreting Eq. (\ref{lgh}) as the first
term in the expansion of a power law given by Eq. (\ref{dim}), we find that
the dimension of the cluster is given by
\be
\label{cldim}
D=gc/v.
\ee

\begin{figure}[!t]
\begin{center}
\leavevmode
\epsfig{figure=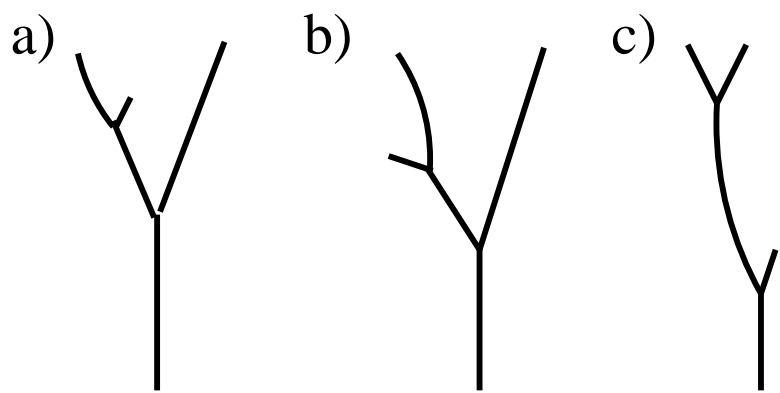,height=8cm,angle=0,scale=.75}
\end{center}
\caption{Competition of three branches.  Various possibilities.}
\label{fig2}
\end{figure}

For $\eta<4$, the number of tip-splitting events proliferates.  The
average number of branches in a system at macroscopic scale $\tilde L$, 
scales as
\be
\tilde L^{\frac{4-\eta}{2}},
\ee
as seen by the lowest term in the $\beta$ function above.
Go to second order in $g^2$ to compute the next term in the $\beta$ function.
If two tip splittings occur, there are now three competing branches, and the
enhanced competition reduces the chance that any two of them will survive
till long times.   
There is a simple physical reason for the enhanced competition.  Eq.
(\ref{pgr}) may seem surprising, as it implies that for $\eta<2$ the 
branches do not compete, while it is known that for any $\eta>0$ a smooth
surface is unstable.  However, the factor $t^{-1}$ in Eq. (\ref{pgr})
is due to the increasing separation of the growth tips over time.  Starting
with perturbations on a smooth surface, there is no such mechanism
causing the perturbations to spread in space.  Similarly, if there
is a large number of competing branches, they are forced closer to each
other and are unable to spread apart from each other as rapidly, enhancing
the competition.  In Fig. \ref{fig2}, we show various possibilities.
In (a), the small
branch in the middle is actually reducing the competition of the two outer
branches, while in (b) and (c) the
competition of the two larger branches is enhanced.  To determine
which of these effects is stronger requires a calculation.

Let the second splitting occur at time $t$ after
the first splitting.  If one of the two branches resulting from the second
dies after a time much less than $t$, the second splitting has no effect.
Thus, the probability that the second splitting will effect the evolution
of the first pair of branches is proportional to $\int {\rm d}t/t$,
again yielding a logarithm as desired.  

Precisely, consider the number of surviving pair of branches at 
scale $\tilde L$ which are produced in time $T$.  If there are no tip splitting
events, this probability is zero; if there is one it is given by the
calculation above.  If there are two, the branches compete and one must
integrate over the initial time at which the two branches are created as
well as over the initial conditions after each tip splitting event.
It is convenient to define $\tilde g$ to be the {\it rate} at which {\it any}
tip splitting event occurs.  Only some fraction of these events lead to
a pair of branches surviving for sufficient time, so that $g<\tilde g \tilde T$.
The number of surviving pairs, to order $\tilde g^2$, is
\be
\label{3b}
P_1 \tilde g T e^{-\tilde g T} + 
\int {\rm d}t P_2(t) \frac{\tilde g^2}{2} T e^{-\tilde g T} + ...=
P_1 \tilde g T - P_1 \tilde g^2 T^2+
\int {\rm d}t \, P_2(t) \frac{\tilde g^2}{2} T+ ...
\ee
where $P_1$ is the probability that one tip splitting event gives rise
to a surviving pair of branches, while $P_2(t)$ is the probability that
two such events produce a surviving pair, with $t$ the time difference
between events.  Note $P_1 \propto l/\tilde L$.

If $t>>\tilde T$, then the branch creation events are independent, and we 
find that $P_2=2 P_1$ as either of the two creation events may give
rise to a branch pair at scale $\tilde L$, so that terms of order $T^2$ cancel 
in the above equation.  Otherwise, $2 P_1-P_2$ is of order $1/|t|$; assume
\be
\label{p2f}
P_2=2 P_1-c_2/|t|,
\ee
for some constant $c_2$.  Then, the bare constant
$g$ is determined by 
\be
g=P_1 \tilde g \tilde T,
\ee
and the probability of producing a surviving branch pair at scale
$\tilde L$ in time $\tilde T$ is given by
\be
\tilde T(P_1 \tilde g - \frac{c_2}{2}
\tilde g^2 \int \frac{{\rm d}t}{t} )
=g-c_2 \frac{g^2}{\tilde T P_1^2} \ln{\tilde L/l}.
\ee
This yields the $\beta$ function, taking into account velocity renormalization,
\be
\label{bfn}
\beta(g)=\frac{\epsilon}{2} g-c_2\frac{g^2}{\tilde T P_1^2}+c\frac{g^2}{v}.
\ee

To evaluate $c_2$, as well as $c$ in Eq. \ref{cldim}, we must turn to
a numerical renormalization group as discussed in the next section,
summing over different possible branch creation events.  
We will numerically evaluate $P_2(t)$ for a fixed $t>>l/v$, and then use the 
scaling arguments given here to obtain $P_2(t)$ for any $t$
and get the lowest order $\beta$-function.  Similarly, $c$ can be
obtained by considering only one set of initial conditions
after the tip splitting event and using scaling arguments.  
The RG requires numerical
work, but remains essentially an analytical treatment, in the same sense
that we may refer to an RG in field theory as analytical despite
possible use of a computer to perform loop integrals.

There is an interesting question, related to the correct form of the
$\beta$ function for various definitions of $g$, whether $g$ is the
rate at which branch pairs are produced that survive for to
scale $\tilde L$ or until time $\tilde L/v$.  The second definition
is what is needed to determine the number of branch pairs in the system.
At time $t$ after the first tip-splitting event, let there be a second
tip splitting event, so that one of the
second pair of branches survives only for a time $t'$ much less than $t$.
Then, when the two remaining branches reach separation $\tilde L$,
a time $\tilde L/v+t'$ has elapsed, rather than a time $\tilde L/v$.
We will find that, depending on which definition of $g$ is used,
$P_2(t)$ will differ by an amount of order $t'/\tilde L$.  Note that
this is independent of $t$, differing from the form we have
assumed in Eq. (\ref{p2f}).  This
yields a change in $\int {\rm d}t P_2(t)$ independent of $L$, and of
order $t'$; integrating over initial conditions for the second branching
event will yield a logarithmic correction to $\int {\rm d}t P_2(t)$.  However,
since the velocity is renormalized, there
is an additional term in the $\beta$ function for the first
definition, equal to $\frac{g^2 c}{v}$, which gives the same
correction to $P_2(t)$.  However, we will not need to consider this
complication, because for fixed $t'$ and $\tilde L$ large, the change
in $P_2(t)$ is negligible and this is the regime in which
we perform the numerical calculations.

Similarly, there may also be slight differences in the definition of
$g$ between requiring that the branch pair grow until {\it height} $\tilde L$
or until {\it separation} $\tilde L$.  
Although at long times the surviving branch
pair grows with the characteristic 72 degree angle, and the height and
separation are related, there can again be deviations in the separation by
an amount of order $1/\tilde L$.  If the present work, concerned with
the stationary regime in the cylindrical geometry, were extended to the affine
regime, perhaps these deviations in separation will be connected with
different scaling in horizontal and vertical directions.
\section{Numerical Renormalization Group}
The reduced model was implemented numerically in a
discrete version, such that the program alternates
between growth tips on different time steps,
advancing each growth tip by an amount proportional to the growth
probability measure.  To solve Laplace's
equation quickly, the method of iterated conformal maps\cite{conf,num2} is
used, using ``strike mappings"\cite{conf} to obtain branches of vanishing
aspect ratio, and computing the Jacobian at the tips by appropriate
regularization.

To obtain dynamics in the scaling regime well below the macroscopic
cutoff, we simply take this cutoff to infinity, which poses no
difficulty in a conformal mapping implementation of the reduced
model: the surface that is
mapped onto the growing cluster is the real line, while the map
$z\rightarrow z^2$ is used to produce a single growing tip as an initial
condition.  To evaluate the $\beta$-function, we run the simulations until
the two most active branches become more than a large, but finite, distance 
$\tilde L$ apart.  Of course, if it were not necessary to perform the 
calculations
numerically, this distance would be taken to be infinite.  
To produce tip splittings, we use a function whose derivative
is given by Eq. (\ref{feq}) to produce a pair of growth tips, choosing
the initial separation between tips to be small.
Interestingly, this function
can be obtained in closed form as (for $x=1$)
\be
\label{cl}
F(z)=z^{2/5} (z^2-5)^{4/5}.
\ee

The strike mappings take a particularly simple form in this geometry.
We have $f(z)=\sqrt{z^2-\lambda}$.  Using the shape of the map $F(z)\propto
F_i (z-z_i)^2$ near the tip, this increases the length of the branch by
$F_i \lambda$.  The regularized value of $F'(z)$ near the tip is
$\sqrt{F_i}$, but $\lambda$ is taken proportional to $|F'|^{-2}$, so
that we produce strikes of constant length in the physical plane.

With this geometry, the program runs significantly faster than such a
program in radial geometry, due to the simple form of the map functions.
It may be worth investigating this as a means of speeding up simulations
using the conformal mapping model.

The errors due to discretization can be surprisingly large, so that
even after thousands of time steps there are noticeable, though small,
deviations from the linearized branch competition dynamics discussed
above.  We have verified that these errors are reduced as the discretization
is reduced, but we have had to make some compromises to obtain a sufficiently
fast program.

To obtain a range of different initial conditions for branches after
splitting, we use the discretized version of the reduced model, but
we introduce a factor $f$ which multiplies the growth velocity of one of
the daughter branches on the {\it first} step after splitting.  When
this factor is close to 0, that daughter tends to lose, when it is close to
1, that daughter tends to win; a balance arose when $f$ is near 
$f_0\approx 0.278$.
We are able to start the branches sufficiently close that they compete
roughly equally for tens of thousands of growth steps.

As discussed above, we fix the time $t$ of the second splitting (chosen
to be 500 steps for each of the two branches), pick one of the
two branches to split at random with relative probability equal to
the relative growth measure on the tips, and then evolve the three branches.
Typically, for the two most active branches to obtain the chosen
separation requires of order another 3000 growth steps per branch.

We wish the initial conditions for the second tip-splitting event to have the 
same microscopic cutoff as for the first tip-splitting event.  In order
to do this, the same mapping of Eq. (\ref{cl}) is used for both tip-splitting
events, with an appropriately scaled value of $x$, and the 
overall velocity scale
after the second tip-splitting event is chosen such that those two daughters
initially have the same step size in the {\it physical plane}
as the two daughters after the first
tip-splitting event.

We tried two methods of handling the integration over initial
conditions.  First, a Monte Carlo technique:
initial conditions are chosen randomly and we count the fraction of initial
conditions which give rise to a surviving branch pair.  To speed the
Monte Carlo, we sampled only a subset of initial conditions, chosen to
include all conditions which could give rise to a surviving pair.
By having the program randomly chose which of the two branches after the second
branching event would get the first growth step, we ensured that the 
distribution of initial conditions for this event be symmetric between the 
branch growing inward and the branch growing outward.  
However, the Monte Carlo
technique suffered from the problem that, to obtain a sufficient number
of surviving branch pairs in the available computer time, we had to adopt a 
fairly broad definition of ``surviving" and keep $\tilde L$ finite.  
However, the correct limit for the renormalization group is to take 
$\tilde L>>l$.  In this limit, it does not matter how the survival of
a branch pair is defined, but for finite $\tilde L$, we more closely
approximate the desired results if we take the definition of
survival to be very narrow, so that the two branches must be almost identical
in growth measure.  This was accomplished by manually searching for initial
conditions which gave rise to a surviving branch pair (with an extremely
narrow definition of survival so that $\delta$ for the two surviving
branches was almost exactly zero) and numerically computing the derivative of 
$\delta$ with respect to the initial conditions to obtain $P_1,P_2$.

If we fix the initial conditions of one of the tip-splitting events to produce
a very asymmetric pair
of daughters so that after a brief time only one daughter survives, and
choose the initial conditions of the other tip-splitting event randomly,
the probability that a branch pair survives to scale $\tilde L$ must approach
$P_1$. 
Due to the discretization errors in our program, this probability
is within a few percent of $P_1$, but not exactly equal to $P_1$;
we have verified that this difference also disappears as the discretization
is reduced.  The numerical evaluation of $P_2-2 P_1$ suffers from
the problem that the large contributions resulting from the very narrow region
where all three branches compete may be swamped by small errors over
the large region of all other initial conditions.  Thus,
we have had to adapt some criterion for determining what range of
initial conditions to integrate over.  This necessarily introduces some
error and makes our results for the $\beta$ function somewhat subjective.
It is hoped that improved numerical techniques will at some point improve
this situation.  One reason for the difficulty is that the processes in Fig.
\ref{fig2} contribute to the $\beta$ function with different signs.  
Another reason for these small errors is discussed at the end of the
last section in terms of corrections of order $1/\tilde L$ which
are logarithmically divergent when integrated over initial conditions
for the second surviving branch pair; perhaps using a different definition
of $g$ will improve the situation.

The constant $c$ in Eq. (\ref{cldim}) is obtained by evaluating
a process with a single tip splitting.  We consider the change in height
of the dominant branch of the pair as compared with the height of a 
branch without splitting.
The simulation of the branch pair was run for approximately
50000 steps per branch.  The change in height fits very well the form
$c-c_1/\sqrt{t}$, with $c_1$ some other constant; using this form we were
able to obtain the asymptotic change in height by extrapolation.
This quantity $c$ can be obtained much more accurately than $c_2$.

Using Eq. (\ref{cldim}) evaluated at the fixed point
of Eq. (\ref{bfn}), I have obtained that 
\be
D\approx 0.46 \epsilon.
\ee
Using different estimates for the range of initial conditions over which
to integrate, I have obtained a lower estimate of $D \approx 0.40 \epsilon$
and an upper estimate of $D\approx 0.51 \epsilon$.
I then compared to the dimensions obtained numerically\cite{num2}.
The RG result seems high, for at $\eta=3$ we have $D\approx 0.26$ 
and at $\eta=3.5$ we have $D\approx 0.16$, where these results are obtained
by subtracting unity from the dimensions in radial geometry.  However, fitting
the dimensions obtained numerically for $\eta=0,1,2,3$ to a polynomial
in $\epsilon$, I found that the lowest order result is
\be
D\approx 0.45 \epsilon,
\ee
in good agreement with the RG result.  I did not include
$\eta=3.5$ in the polynomial fit as this dimension is known
less accurately than the others and has an inordinate effect on
the lowest order term in $\epsilon$; taking $D=0.17$ at $\eta=3.5$ leads to
$D\approx 0.48 \epsilon$, while taking $D=0.16$ leads to $D\approx
0.41 \epsilon$.

Let us now consider the multifractal spectrum\cite{dq,tq}.
Let $x$ be a point in the physical plane, with
$z=F^{-1}(x)$.  The harmonic measure is $|F'(z)|^{-1}{\rm d}x={\rm d}z$.
Define $p(x)$ to be the {\it normalized} harmonic measure 
\be
\label{pdef}
p(x){\rm d}x=\frac{|F'(z)|^{-1} {\rm d}x}{\int \frac{{\rm d}x'}{l}
|F'(z')|^{-1}},
\ee
and the exponents $\tau(q)$ by
\be
\label{tqd}
\langle \int \frac{{\rm d}x}{l} p^q(x) \rangle
=\Bigl( \frac{l}{L}\Bigr)^{\tau(q)},
\ee
where the angle brackets denote averaging over realizations of the
cluster. 
For an isolated branch in cylindrical geometry, if we pick a
parametrization such that $F(z)\propto z^2$ near
the tip, then the denominator of Eq. (\ref{pdef}) is of order $\sqrt{L}/l$.
For $q>2$ the integral of Eq. (\ref{tqd}) is divergent near the tip and is 
cutoff at 
length scale $l$.  For $0<q<2$ this integral is divergent away from the tip,
and is cutoff at a physical length scale $L$, when the power law behavior of the
electric field crosses over to an exponential decay.  For $q\leq 0$, these
exponents are ill-defined in this geometry.
For an isolated branch, we have 
$\tau(q)=q/2$ for $q>2$ and $\tau(q)=q-1$ for $q<2$.

It is also possible to compute corrections to the multifractal spectrum
in the RG.  Various multifractal exponents may be defined, including
quenched and annealed exponents\cite{qa}.  It has been argued\cite{qa2}
that asymptotically all these exponents are identical for DLA, but that
for finite size systems the apparent exponents may be quite different.
The different exponents all involve different ways of choosing how to
weight different realizations of the cluster when performing the
average in Eq. (\ref{tqd}).
Within the RG framework, we need to have a multiplicatively renormalizable
operator in order to obtain power laws from the lowest order computation.
This requires choosing the correct set of exponents and the correct average,
as will be discussed below.

Let us first consider corrections to $\tau(q)$ for $q>2$.
The integral of Eq. (\ref{tqd}) must be summed over
all configurations of the cluster.  The configuration with only one
branch gives the zeroth order result.  
Let $\int p^q$ for the configuration with a single
branch be equal to $P_0(q)$.  
At first order in $g$ one must sum over 
configurations with two branches also. 
One must integrate $\int p^q-P_0$
over the full trajectory of the two branches, starting with the two
branches almost equal and continuing until one branch has completely won out.
This trajectory can be obtained numerically.
Note that the integral of $p^q(z)$ will be dominated by the field near the tips,
and so it suffices to know $F_i$ at each tip:
\be
\int \frac{{\rm d}x}{l} p^q(x) \propto 
\Bigl( \frac{l}{L} \Bigr)^{q/2}  \sum\limits_{i} F_i^{-q/2}.
\ee

Suppose the integral over the branch trajectory of 
$\int p^q - P_0(q)$ is equal to $-t x P_0$,
where $t$ is the time the two branches live.
Integrating over initial conditions of the two branches,
we have that the left-hand side of Eq. (\ref{tqd}) is equal to
\be
\label{rsm}
\Bigl( \frac{l}{L}\Bigr)^{q/2}
\Bigl(P_0- x g P_0 \ln{L/l} + ...\Bigr)
\ee
giving $\tau_{\eta}(q)=q/2 + x g$.

How should we perform the integral over the branch trajectory?  Each
configuration along the trajectory corresponds to a different realization 
of the cluster, and depending on how we choose to weight different
realizations, we must weight the integral over the trajectory differently.
We choose to weight the integral by 
\be
\Bigl(\sum \limits_{i} \sqrt{F_i}^{-\eta}\Bigr)^{-1} {\rm d}t,
\ee
which corresponds to changing the time scale such that each branch
grows at a rate proportional to $\sqrt{F_i}^{-\eta}$, rather than that given
by Eq. (\ref{reggm}).  However, this is exactly what happens with a
large number of branches if one of the branches splits: in this case, 
the denominator of Eq. (\ref{reggm}) involves contributions from all 
branches and is insensitive to the splitting on the single branch, so
that we expect that this choice of weighting function will define
$\langle \int p^q \rangle$ in a manner that is multiplicatively
renormalizable.

Further, with this choice, the dimensions will automatically 
obey the electrostatic scaling law\cite{elect} order by order.  
This scaling law adapted to cylindrical geometry is
\be
1+D=\tau(2+\eta)-\tau(\eta).
\ee
The electrostatic law is obtained, in this geometry, by noting
that the growth velocity is proportional to the average,
using the growth probability measure,
of $\lambda$ over tips: this average scales
as $\tau(2+\eta)-\tau(\eta)$.  Since this relation between the growth
velocity and the average of $\lambda$ holds for each branch configuration
separately, it holds for the exponents as we have defined the average
above.

We have obtained the lowest order results that
\be
\tau(6)=3+1.77 D, \tau(5)=2.5+1.35 D,\tau(4)=2+.77 D,\tau(3)=1.5-.23 D.
\ee
We report these results in terms of $D$ by using Eqs. (\ref{cldim},\ref{rsm})
as these calculations are more accurate than that for Eq. (\ref{bfn}).
Note that even at lowest order in $\epsilon$ the $\tau(q)$ spectrum
does not have any simple form such as a log-normal distribution or gap
scaling.
By looking at larger $q$, we find that the Turkevich-Scher\cite{ts} scaling 
law does not hold
as an equality in this expansion.  It holds only as an inequality.

For $0<q<2$, a similar calculation can be performed, although it will be
slightly more complicated as the integral of $p^q(z)$ is no longer
dominated by the tip and must be considered over the full branch.  Further,
the cutoff $L$ of cylindrical geometry must be kept finite, rather
than being taken infinite as we have done in the calculations above,
to keep the integral of $p^q$ finite.  

One further complication is
that $\tau(q)$ is not an analytic function of $4-\eta$ near $q=2$.
The $\tau(q)$ above correspond to an $f(\alpha)$
spectrum with two points: $f=.5,\alpha=0$, the zero
dimensional set of singularities near the tips, and $f=1,\alpha=1$, the
rest of the branch forming a one dimensional set.  As $\eta$
decreases, the singularities near the tips soften while the dimension of
the set of tips increases, and the function $f(\alpha)$ becomes a curve
rather than a set of discrete points.
The function $\tau(q)$ above is not analytic in $q$ as for
$q>2$ the integral of $p^q$ is dominated by $\alpha=0$,
while for $q<2$ it is dominated by $\alpha=1$.
Similarly, for fixed $q$ near $2$, as $\eta$ is decreased the value of 
$\alpha$ 
which controls $\tau(q)$ may jump discontinuously from near 0 to near
1, or vice versa.  It may be possible to surmount this problem by expanding the
left-hand side of Eq. (\ref{tqd}) in $q-2$ and trying to resolve the
logarithms that result as a sum of different power laws.
\section{Renormalizability}
Clearly, it will be very difficult to prove renormalizability, since
we find it difficult to evaluate even lowest order processes, but
renormalizability is essential for the RG we employ.  Further, if we prove
multiplicative renormalizability for the theory with two parameters, $v,g$,
then the fractal nature of DBM cluster will follow without any detailed 
calculation of $\beta$-functions: it
is clear that $v$ has no effect on the RG flow of $g$, and if
$g$ has an attractive fixed point, then
without fine-tuning the theory must arrive at a critical point with nontrivial
exponents.  If instead $g$ had a repulsive fixed point, we would observe
very different critical behavior: for $\eta>4$, there would be a phase
transition in the behavior of the system as a function of the bare
coupling constant from one-dimensional to fractal; the bare coupling
constant could perhaps be adjusted using a noise reduced version of the
model.  Also, for fixed bare $g$, there would be a discontinuous
change in the dimension of the cluster as a function of $\eta$ from $D>0$ to
$D=0$.
So, let us sketch how such a proof might proceed.  

Consider a collection of competing branches with separation distances
of order $L_1$.  Let one branch have single tip-splitting event such
that after a time $T_2$ only one of the two subbranches survives, with
$v T_2 << L_1$.  While both subbranches survive the total growth
measure on the pair is reduced compared to that on the subbranch, but
on times much greater than $T_2$, the growth measure of the surviving
subbranch asymptotically approaches that of the parent, and
at scale $L_1$ it seems simply as if the branch with the tip-splitting
event had grown with a renormalized velocity for a 
brief time.  After integrating over $T_2$, the renormalization of the velocity 
is of order $\ln{L_1/l}$.  

It would be disastrous if instead there were a logarithmic
divergence which depended  on $L_1$, such as $L_1 \ln{L_1/l}$, as
this would change the growth rules of the theory and the manner in which
branches compete.  However, since we are interested in the divergence
as $T_2 \rightarrow 0$, we can restrict to $v T_2<L_1$ and formally expand
the growth velocity in
a power series of $v T_2/L_1$ (this is possible since, for $T_2$ small,
the influence of the branch pair on the other branches is small), 
so that the growth velocity is
\be
\sum\limits_{k} \int \frac{{\rm d}T_2}{T_2} a_k (v T_2/L_1)^k.
\ee
The zeroth order term in this series yields a logarithmic divergence of
desired form, while the higher terms are not divergent.

It would equally be disastrous if the growth measure of the surviving
subbranch were not to asymptotically approach that of the parent.  However,
this follows from properties of Laplace's equation: at long times the
surviving tip is at a height much greater than the dead subbranch,
and the dead subbranch has only a small effect on the growth probabilities
near the surviving tip.

This show that processes with a single branching just renormalize
$v$.  Consider several branchings.  Each set of branchings is defined
by a tree diagram indicating the topology of the branchings, as well as
a set of times $t$ indicating the times at which branching take place
as well as a set of initial conditions for each branch pair.
To the initial conditions corresponds another set of times,
the times that the two branches live.  For fixed topology,
one must integrate over these times and evaluate the average rate at
which the cluster grows or produces branches.  
Let us order the times from smallest to largest, and send groups of these
times to zero.  If a set of times $t_i$ are sent to zero together, with some
other times $t_a$ being held fixed, one finds, as above, logarithmically
divergent renormalizations of $v$ and $g$.  However, by expanding in
$t_i/t_a$ as above, we again obtain the desired result that the
logarithmically divergent terms depend only on a log of the ratio of $t_a/t_i$.

These arguments sketch the renormalizability of the theory.  One
must also show, for example, that there are not divergent contributions
to the rate at which one branch splits into three.  

Also, in defining
the reduced model, we ignored shot noise in the velocity of a given branch;
however, the process of tip splitting, which renormalizes the average
velocity, will reintroduce some noise in the velocity.  The noise in
the velocity of a branch is irrelevant at long time; fortunately, then,
while the renormalization of velocity is logarithmically divergent, the
rms fluctuations in the velocity are convergent at short
times, behaving as (taking
the branch pair to survive for time $t$) $\int \frac{{\rm d}t}{t^2} {t^2}$.

In the continuous model,
$\delta$ increases in time, and the probability of having $\delta=0$ decreases
in time, such that this probability, multiplied by the branch length,
remains constant.  Consider now the case of a discrete formulation of the
model, to see the effects of shot noise or other
fluctuations in the velocity.  Suppose two competing branches
have lengths $1+\delta$,$1-\delta$.  Take a discrete growth step, adding
length $2 \delta$ to one of the branches, with probabilities 
$1/2 \pm \delta$.  The probability of having the two branches symmetrically
distributed after the discrete growth step is $1/2-\delta$.  If the
initial configuration of $\delta$ is chosen uniformly near $\delta=0$, we
find that the probability of having $\delta=0$ is again decreased with
this probability, multiplied by branch length, again remaining constant
to linear order in $\delta$.  To second order in $\delta$ we will find
differences between the continuous and discrete models.

One can do a similar calculation for the case in which the growth of
a branch stagnates due to a tip splitting event.  The probability of
a given branch having a tip splitting event is $1/2\pm\delta$, while
the tip splitting event will slow that branch for a time $1/2\mp\delta$
over during which the other branch grows at a velocity $1/2\mp\delta$.
Again one can show that the probability of having $\delta=0$, multiplied
by branch length, remains constant, to linear order in the time of the
tip splitting event.

This comparison between continuous and discrete models is why
we emphasize that the rms fluctuations induced by
tip-splitting are convergent.  Suppose
instead that the rms fluctuations were not convergent.  Then, one would
find in Fig. 2a,b that $c_2$ was divergent when integrating over
initial conditions for the second branching event, which would lead
to divergences of the form $(\ln{L/l})^2$ destroying renormalizability.
\section{Conclusion}
We have presented the lowest order in an $\epsilon=4-\eta$ expansion, obtaining
good agreement with numerical results.  For more accurate comparison with
DLA, for which $\epsilon=3$, extension to higher order with
further branching processes is necessary.  Due to
the need for numerical techniques to evaluate even lowest order
processes, it is unclear if higher order terms can be computed accurately.

It is also necessary to improve the numerical evaluation of lowest
order processes.  To numerically simulate the reduced model, some
discretization and cutoff is necessary.  We have used a particular
cutoff, but perhaps other cutoffs are preferable.  Indeed, the discrete
random walker formulation of the DBM provides another cutoff which may
lead to more accurate results; one can simulate the DBM with growth confined
to occur on only two or three different tips to generate a system with
a fixed number of branches.

It is also interesting to consider the extension to higher dimensions.
While conformal mapping techniques were very useful in all the calculations
above, the basic idea, determining the relevance of tip-splitting based
on the competition of two branches, should be applicable in any dimension.
While there is a lower bound on the dimension of DLA in higher 
dimensions\cite{bbnd}, there does not seem to be any such bound for the DBM
which would prevent the existence of an upper critical $\eta$.
\section{Acknowledgements}
I would like especially to thank T. C. Halsey for explaining his
branched growth model and for sharing his unpublished idea
of systematically improving the branched growth model by a ``fugacity 
expansion" in the tip splitting rate.
I would also like to thank L. Levitov, F. Levyraz, P. Pfeifer,
and I. Procaccia for useful discussions, as well as the Centro Internacional 
de Ciencas in Cuernavaca for a very interesting recent conference on
DLA and other nonlinear systems.  This work was supported by DOE grant 
W-7405-ENG-36.


\begin{thebibliography}{99}
\bibitem{dla}  T. A. Witten and L. M. Sander, Phys. Rev. Lett. {\bf 47}, 1400
(1981).

\bibitem{vf} J. Nittman, G. Daccord, and H. E. Stanley,
Nature {\bf 314}, 141 (1985).

\bibitem{dep} D. Grier, E. Ben-Jacob, R. Clarke, and L. M. Sander, 
Phys. Rev. Lett. {\bf 56}, 1264 (1986); R. M. Brady and R. C. Ball, Nature 
{\bf 309}, 225 (1984).

\bibitem{dend} J. Kert\'esz and T. Vicsek, J. Phys. A {\bf 19}, L257 (1986).
 
\bibitem{thy} B. Davidovitch, A. Levermann, and I. Procaccia, Phys. Rev. E 
{\bf 62}, R5919 (2000).

\bibitem{thy2} T. C. Halsey and M. Leibig,
Phys. Rev. A {\bf 46}, 7793 (1992);
T. C. Halsey, Phys. Rev. Lett {\bf72}, 1228 (1994).

\bibitem{thy3} A. Erzan, L. Pietronero, and A. Vespignani, Rev. Mod. Phys. 
{\bf 67}, 545 (1995).

\bibitem{ft} G. Parisi and Y. C. Zhang, J. Stat. Phys. {\bf 41}, 1 (1985);
L. Peliti, J. Phys. (Paris) {\bf 46}, 1469 (1985).

\bibitem{dbm}  L. Niemeyer, L. Pietronero, and H. J. Wiesmann, Phys. Rev. Lett.
{\bf 52}, 1033 (1984).

\bibitem{eta0} M. B. Hastings, preprint cond-mat/9910274.

\bibitem{num1} A. Sanchez et.~al., Phys. Rev. E {\bf 48}, 1296 (1993).

\bibitem{num2} M. B. Hastings, preprint cond-mat/0103312.

\bibitem{slg} M. A. Peterson, Phys. Rev. E {\bf 57}, 3221 (1998).

\bibitem{unpub} T. C. Halsey, private communication.

\bibitem{ng} N. Goldenfeld, Lectures on Phase Transitions and the
Renormalization Group (Addison-Wesley, New York, 1992).

\bibitem{diffd} P. Meakin and S. Tolman, in Fractal's Physical Origin
and Properties, Ed. L. Pietronero (Plenum, New York, 1989); 
see also Ref.~\cite{thy3}.

\bibitem{evert} C. Evertsz,
Phys. Rev. B {\bf 41}, 1830 (1990).

\bibitem{conf} M. B. Hastings and L. S. Levitov, Physica D {\bf 116}, 244
(1998).

\bibitem{clg} F. Barra et. al., preprint cond-mat/0103126.

\bibitem{dq} H. G. E. Hentschel and I. Procaccia, Physica D {\bf 8}, 435 (1983).

\bibitem{tq} T. C. Halsey et. al., Phys. Rev. A {\bf 33}, 1141 (1986).

\bibitem{qa} M. E. Cates and T. A. Witten, Phys. Rev. A {\bf 35}, 1809 (1987).

\bibitem{qa2} T. C. Halsey, B. Duplantier, and K. Honda, Phys. Rev. Lett.
{\bf 78}, 1719 (1997).

\bibitem{elect} T. C. Halsey, Phys. Rev. Lett. {\bf 59}, 2067 (1987).

\bibitem{ts} L. A. Turkevich and H. Scher, Phys. Rev. Lett. {\bf 55}, 1026
(1985); T. C. Halsey, Phys. Rev. A {\bf 38}, 4789 (1988).

\bibitem{bbnd} R. C. Ball and T. A. Witten, Phys. Rev.
A {\bf 29}, 2966 (1984).
\end{thebibliography}
\end{document}